\documentstyle[pra,aps,preprint]{revtex}
\begin{document}
\title{Frequency dependent hyperpolarizabilities of atoms;
calculations using density-functional theory}
\author{\bf Arup Banerjee and Manoj K. Harbola}
\address{Laser Physics Division
Center For Advanced Technology, Indore 452013, India}
\maketitle
\begin{abstract}
Using the orbitals generated by the van Leeuwen-Baerends potential [Phys.
Rev. A {\bf 49}, 2421 (1994)], we calculate frequency-dependent
response properties of the noble gas atoms of helium, neon and argon
and the alkaline earth atoms of berrylium and magnisium, with particular
emphasis on their nonlinear polarizabilities.  For this, we employ the
time-dependent Kohn-Sham formalism with the adiabatic local-density
approximation (ALDA) for the exchange and correlation.  We show that 
the results thus obtained for  
frequency-dependent polarizabilities (both linear and nonlinear) 
of the inert gas atoms are highly accurate. On the other hand,  
polarizabilities of the alkaline earths are not given with the
same degeree of accuracy. 
In light of this, we make an assessment of ALDA for 
obtaining linear and nonlinear response properties by employing 
time-dependent density-functional theory.
\end{abstract}
\newpage
\section{Introduction}
Over the past decade, the time dependent density-functional theory 
(TDDFT) \cite{gross1,casida} is being used increasingly to study frequency 
dependent response properties
\cite{zangwill,stott,ghosh,barto1,senatore,zong,gisberg1,baner1,gisberg2,gisberg3,baner2,aiga}
and excitation energies \cite{casida,gisberg2,petersilka} of many-electron 
systems. The theory found its first application in the calculation of 
frequency dependent polarizabilities of
atoms and photo-absorption cross section of atoms and molecules 
\cite{zangwill,stott}. However, the existence of TDDFT was not formally 
proved at that time. It was some years later that Deb and Ghosh \cite{deb1},
Bartolotti \cite{barto2} and Runge and Gross \cite{runge} 
laid rigorous foundations of TDDFT. Analogous to its stationary counterpart
TDDFT is exact in principle but its practical implementation requires
approximating the exchange-correlation (XC) energy functional. In most of 
the applications the adiabatic local density approximation (ALDA) 
\cite{gross1} is used. In this approximation the nonlocal time-dependence of 
the functional is ignored and the spatial dependence is also treated locally. 
As such the functional has the same form as the local-density approximation 
(LDA) of the stationary-state theory \cite{parr,dreizler}. Results for the 
response properties using the LDA for the unperturbed system and ALDA for 
perturbation calculations show that with these approximations the 
polarizability at zero frequency (static polarizability) as well as its 
frequency dependence are overestimated. Use of LDA+ALDA for the response 
property calculations introduces errors at two different levels.  
First, the use of the LDA for calculation of the unperturbed orbitals gives 
potentials and densities which are asymptotically not correct. In particular 
the tail of the LDA potential is exponential rather than its correct 
$-\frac{1}{r}$ behaviour \cite{almb}. Thus the physical propreties that 
depend on the 
asymptotic nature of the ground-state orbitals and densities, for example 
the response properties, are not determined accurately by the LDA orbitals. 
Secondly, errors in the response property calculations arise beacause: 
(i) further change in the potential itself is being calculated approximately,
and (ii) ALDA is valid only in the limit of zero frequency. There have been
attempts to rectify the problem at the first level by employing 
methods which reproduce the asymptotic behaviour of the potential correctly.
Thus Senatore and Subbaswamy \cite{senatore} applied the self-interaction 
correction \cite{perdew} method to obtain accurate orbitals with proper 
asymptotic decay. Later Zong et. al \cite{zong} devised a scissors operator
technique and Gisbergen et al. \cite{gisberg1} employed a model potential with 
the desired asymptotic behaviour to improve the polarizabilities. Gisbergen 
et al. \cite{gisberg2} also studied frequency dependent polarizabilities of 
He, Be and Ne by employing exact XC potential for the ground-state of these 
atoms coupled with ALDA for the XC kernels. Nearly exact ground-state 
wavefunction along with the ALDA for XC kernels were employed by us 
\cite{baner2} to calculate nonlinear optical coefficients of helium, and to 
investigate the accuracy of ALDA in predicting the frequency dependence of 
nonlinear polarizabilities.

All these studies show that the inaccuracy in the response properties
arise mainly from the use of the LDA to obtain the unperturbed densities;
For perturbative calculations, on the other hand, ALDA appears to be a
reasonably accurate approximation, particularly at frequencies in the 
optical range.  One may therefore conclude that if the asymptotic nature
of the potential is corrected, the response properties at optical frequencies
will come out to be copmparable to the ab-initio or experimental values
even with the use of ALDA.  However, the conclusions above cannot
said to be general since most of the studies have been confined to 
investigating the linear polarizabilities.

Against this background it then becomes necessary to investigate how does 
ALDA perform in the calculation of frequency dependent nonlinear optical 
coefficients of heavier atoms when asymptotically accurate ground-state 
orbitals are used for this purpose. For the helium atom we could perform such 
a study with a near exact ground-state orbital \cite{koga} obtained from its 
Hylleraas wavefunction. However, this is not possible for systems with larger 
number of electrons. For such systems we generate the ground-state 
orbitals and potentials by using the model potential introduced by van 
Leeuwen and Baerends (LB) \cite{van} as a correction to the LDA XC potential.
Recently we have employed this potential to calculate the static 
nonlinear response properties of several atoms and ions \cite{baner3}. We 
found that the orbitals given by this model potential, when used with the LDA 
for the higher derivatives of XC potential, give reasonably accurate static
hyperpolarizabilities for the inert gas atoms.  Thus to assess the accuracy 
of ALDA in predicting the frequency dependence of nonlinear polarizabilities,
in this paper we first study dynamic hyperpolarizabilities of these atoms
calculated by employing the orbitals generated by the LB potential.
In our study \cite{baner2} of the static properties, we also found that the 
hyperpolarizabilities for the alkaline earths show some improvement
with the inclusion of the LB correction although their linear 
polarizabilities remain unaffected.  In this paper we study the 
frequency-dependent polarizabilities of the alkaline earth atoms of Be 
and Mg also and show that the dynamic polarizabilities of these systems 
follow the same trend as their static counterparts.
In the following, we first briefly describe the LB potential.  We follow
that with a short description of the variation-perturbation method of 
calculating the dynamic response properties.  We then present our results and
conclude the paper with a discussion.

van Leeuwen and Baerends \cite{van} proposed a correction to the LDA potential
primarily to correct its asymptotic behaviour.  For this they first noted 
that the Becke's correction \cite{becke} to the LDA functional gives the 
correct exchange-energy density but fails to give the potential correctly.  
Thus they suggested that a Becke-like correction be added to the potential 
directly, and found that such a correction brings the approximate potential 
very close to the exact one.  Thus the highest occupied orbital eigenenergy 
as obtained from this potential is close to the ionization energy of a 
many-electron system.  More recently, it has been shown that the corresponding
total energy is also quite accurate \cite{baner3} .  The effective potential 
proposed by LB has the form
\begin{equation}
v_{xc}({\bf r}) = v_{xc}^{LDA}({\bf r}) + v^{LB}({\bf r})
\label{1}
\end{equation}
with
\begin{equation}
v^{LB}({\bf r}) = -\beta\frac{\rho^{\frac{1}{3}}x^{2}}{1 + 3\beta x\sinh^{-1}(x)},
\label{2}
\end{equation} 
where $x = \frac{|{\bf \nabla}\rho|}{\rho^{\frac{4}{3}}}$ and $\beta = 0.05$.
Note that the extra term $v^{LB}({\bf r})$ added to the LDA
potential is like the Becke term. It therefore represents the correction to 
only the exchange component of the potential.  
Using the potential given by Eq.(\ref{1}) and (\ref{2}) we generate the 
ground-state orbitals and potentials for the above mentioned atoms and
then employ these orbitals for the calculation of frequency dependent 
linear and nonlinear polarizabilities.

We perform calculations of the optical response properties by employing 
the variation-perturbation method \cite{baner1} within the time dependent 
Kohn-Sham (TDKS) formalism of TDDFT. This method has been used in the past
to calculate \cite{baner2} polarizabilities correponding to several nonlinear
optical phenomena. The method and expressions for various 
hyperpolarizabilities have been discussed \cite{baner1,baner2} in our earlier
works. Thus we do not describe these in detail here.  It is sufficient to 
mention that the linear polarizability and the coefficient corresponding to 
the degenerate-four-wave-mixing (DFWM) are obtained directly by minimizing 
the second-order and fourth-order changes in the quasi-energy with respect
to the first and second-order orbitals, respectively. The other nonlinear 
coefficients are not related directly with the fourth-order energy change.  
However, expressions for these coefficients in terms of second-order orbitals
have been derived \cite{baner2,baner4} within TDKS theory and it 
is these expressions which we employ here.  To perform our calculations, we 
represent the radial part of the induced orbitals by a linear combination 
of the Slater type orbitals (STO). Thus it is given as           
\begin{equation}
f(r) = \sum_{i}C_{i}r^{n_{i}}e^{-\eta_{i}r},
\label{3}
\end{equation}
where $n_{i}$ and $\eta_{i}$ are the parameters of STO which are fixed and
$C_{i}$ are the linear variational parameters.  Parameters $n_{i}$ and
$\eta_{i}$ are chosen in such a way that the functions correctly represent
the excited states. We have optimized these parameters by calculating the
static poalrizabilities and hyperpolarizabilities from the LDA ground-state
orbitals, and matching the results with the numbers obtained by Stott and 
Zaremba \cite{stott}.  For the exchange-correlation energy we have used the 
Gunnarsson-Lundquist parametrization \cite{gl}, which is the same as used by 
Stott and Zaremba. Thus our results for hyperpolarizabilities are around 
10\% less than those of Senatore and Subbaswamy \cite{senatore} who use the 
Perdew-Zunger \cite{perdew} parametrization.

We now present the results of our calculation. Although our main focus in
this paper is on the frequency dependent hyperpolarizabilities, 
for completeness we first discuss the results for frequency dependent
polarizabilities.

In Figs.1-3 we show the ratio $\frac{\alpha(\omega)}{\alpha(0)}$ 
for the inert gas atoms He, Ne and Ar as a function
of frequency $\omega$ and for comparison also display the corresponding 
ab-initio results. For helium we compare our results with the results of 
Bishop and Lam \cite{bishop} and for neon and argon the comparison is made 
with the MP2 results of Rice \cite{rice}.  Since $\alpha(0)$ is already 
reproduced \cite{baner3} quite accurately with the LB orbitals, it is clear 
from these figures that $\alpha(\omega)$ is also accurate and matches quite 
well with the ab-initio results. For helium the ab-initio results are 
essentially reproduced by our calculations for frequencies up to 0.5 a.u. 
(wavelength of about 914 \AA). For neon the match with the MP2 results is 
good till about 0.1 a.u. and for argon good match is obtained up to 
$\omega = 0.07$ a.u.. For comparison we also show the corresponding LDA
results.  It is quite clear that the LDA results are highly inflated in 
comparison to both the LB corrected and the ab-initio numbers.

In Figs. 4 and 5, we present the results of 
$\frac{\alpha(\omega)}{\alpha(0)}$ for the Be and Mg atoms.  The results
for these atoms are quite different from those of the inert gas atoms
discussed above.  Here the LDA and the LB corrected results are
essentially the same.  This is intriguing since the energies and the
highest occupied eigenvalues obtained from the two schemes differ
significantly.  A possible reason for this could be that since the outer
electrons in these systems are less tightly bound, the exchange and 
correlation effects play a relatively more important role in determining
the polarizabilities, and ALDA is not sufficient to represent their
effects accurately. Thus although the unperturbed orbitals are improved
by inclusion of the LB correction, this alone is not sufficient 
to get over the inaccuracy of ALDA for these systems. 

Having discussed linear polarizabilities, we now proceed on to present the 
results for the nonlinear coefficients corresponding to third harmonic 
generation (THG) and DFWM of these atoms. As is the case with 
polarizabilities, these results are also obtained from the orbitals generated
by the LB potential with the use of ALDA for perturbative calculations.

First, we discuss the results for helium. Exact results 
corresponding to the above mentioned nonlinear optical effects for
helium are available over a range of frequencies. In Figs. 6 and 7 we 
display our results in comparison with the ab-initio results \cite{bishop1}.
Plotted in Fig. 6 is the DFWM coefficient $\frac{\gamma(\omega)}{\gamma(0)}$,
and in Fig. 7 we plot the THG coefficient $\frac{\gamma(3\omega)}{\gamma(0)}$ 
as a function of $\omega$. It is evident from the figures that the results
for hyperpolarizabilities obtained from the LB orbitals are also highly 
accurate.  The LDA, on the other hand, overestimates the frequency
dependence of these quantities by a large amount.  A comparative
study like this is not possible at all frequencies for neon and argon
because of the lack of available data.  However, some experimental
results are available \cite{lehmeier,new} at few discrete frequencies for 
all three atoms. We now compare our results with these experimental numbers.

In Table I we present our results for the THG coefficients of 
the atoms considered in this paper at two distinct wavelengths, namely, 
$\lambda = 10550$\AA \ ($\omega\approx 0.0433$ a.u.) and 
$\lambda = 6943$\AA \ ($\omega\approx 0.0658$ a.u.).  We do so because 
experimental results for THG by helium, neon and argon exist at these 
wavelengths.  For a complete picture, we give the results obtained from both 
the LDA and the LB orbitals. It is again clear that although the LDA orbitals 
give a large error in the estimates of these quantities for the noble gas 
atoms, the LB corrections to the LDA eliminates almost all of this error. 
Further, the numbers obtained from the LB orbitals lie within the experimental
error bounds.  For the other two atoms also the LB corrected results for
the THG coefficients are less than the LDA numbers.  However, there are no 
experimental numbers available for these systems.  Nonetheless, on the basis
of their zero frequency results, we expect the LB numbers to be closer
to experiments than the LDA numbers.  In Table II we present the results of 
DFWM coeeficients at the same frequencies, although no experimental data 
exists for this effect. However, we expect these results to be quite accurate
for the noble gas atoms, and moderately accurate for the alkaline earths. 
This is because the maximum deviation from the ab-initio or the experimental 
results is observed for THG coefficients which have already been shown to be
accurate.

To conclude, our study above indicates that when accurate orbitals are 
employed to calculate response properties, ALDA reproduces the frequency 
dependence of linear as well as nonlinear response properties of the noble 
gas atoms quite accurately.  On the other hand, for the alkaline earths its 
behaviour with respect to the linear and nonlinear polarizabilities is quite
different.  Thus it appears that for systems, such as Be and Mg, where the 
electrons are loosely bound, ALDA is not a good approximation to 
calculate the effects of exchange and correlation. In a recent study, it has 
been shown that ALDA is a major component \cite{amico}of the 
exchange-correlation functional for time-dependent hamiltonians.  This has 
been done by considering two electrons moving in a time-dependent potential. 
However, our study indicates that this is true only if the electrons are 
tightly bound.

\newpage
\subsection{Table Captions}
{\bf Table I}: THG coefficients
with and without the LB correction to the LDA along with their
experimental values (in atomic units).

{\bf Table II}: DFWM coefficients
with and without the LB correction to the LDA (in atomic units).  
\newpage

\subsection*{Table I}

\tabcolsep=0.1in
\begin{center}
\begin{tabular}{|c|c|c|c|c|c|c|}\hline
Atom & \multicolumn{3}{c|}{$\lambda = 6943$\AA} &
\multicolumn{3}{c|}{$\lambda = 10550$\AA}  \\ \cline{2-7}
& LDA & LDA+LB & Expt. \cite{new} & LDA & LDA+LB & Expt. \cite{lehmeier} \\
\hline

He & 98.46 & 48.35 & 53.6$\pm 7$ & 89.12 & 45.11 & 44.07$\pm 4.8$ \\
    &       &       & 53.6$\pm 2.4$&      &        &              \\
Ne & 230.30 & 103.68 & 119$\pm 13$ & 202.67 & 95.50 & 78.27$\pm 8.3$ \\
    &        &       & 96.5$\pm 4.8$&       &       &            \\ 
Ar & 2532.8 & 1560.3 & 1691$\pm 167$ & 1925 & 1275.1 & 1021.3$\pm 107$ \\
   &       &       &  786$\pm 48$     &      &      &     \\
   & & & & & & \\
Be & 6.42$\times$ 10$^{8}$ & 1.45$\times$ 10$^{8}$ & - & 1.41$\times$ 10$^{5}$
   & 1.07$\times$ 10$^{5}$ & - \\
Mg & no minimum & no minimum & - & 1.88$\times$ 10$^{11}$
   & 3.17$\times$ 10$^{9}$ & - \\ 
\hline
\end{tabular}
\end{center}

\subsection*{Table II}
\tabcolsep=0.1in
\begin{center}
\begin{tabular}{|c|c|c|c|c|}\hline
Atom & \multicolumn{2}{c|}{$\lambda = 6943$\AA} &
\multicolumn{2}{c|}{$\lambda = 10550$\AA}  \\ \cline{2-5}
& LDA & LDA+LB  & LDA & LDA+LB  \\
\hline

He & 87.62 & 44.57 &  84.86 & 43.58  \\
Ne & 198.16& 94.06 & 190.45 & 91.65  \\
Ar & 1863.0 & 1226.1 & 1725.5 & 1154.8 \\
   & & & & \\
Be & 9.83$\times$ 10$^{4}$ & 7.33$\times$ 10$^{4}$ &
   6.17$\times$ 10$^{4}$ & 4.83$\times$ 10$^{4}$ \\
Mg & 4.46$\times$ 10$^{5}$ & 1.98$\times$ 10$^{5}$ &
   2.08$\times$ 10$^{5}$ & 1.14$\times$ 10$^{5}$ \\
\hline
\end{tabular}
\end{center}
\newpage
\subsection*{Figure Captions}
{\bf Fig. 1}: Plot of $\alpha(\omega)/\alpha(0)$ as a function of 
frequency $\omega$ for helium.  The squares, open circles
and filled circles represent the LDA, the LDA$+$LB and ab-initio results
\cite{bishop}, respectively.

{\bf Fig. 2}: Plot of $\alpha(\omega)/\alpha(0)$ as a function of 
frequency $\omega$ for neon.  The squares, open circles
and filled circles represent the LDA, the LDA$+$LB and ab-initio results
\cite{rice}, respectively.

{\bf Fig. 3}: Plot of $\alpha(\omega)/\alpha(0)$ as a function of 
frequency $\omega$ for argon.  The squares, open circles
and filled circles represent the LDA, the LDA$+$LB and ab-initio results
\cite{rice}, respectively.

{\bf Fig. 4}: Plot of $\alpha(\omega)/\alpha(0)$ as a function of 
frequency $\omega$ for Be.  The squares and filled circles represent 
the LDA and the LDA$+$LB results, respectively.

{\bf Fig. 5}: Plot of $\alpha(\omega)/\alpha(0)$ as a function of 
frequency $\omega$ for Mg.  The squares and filled circles represent 
the LDA and the LDA$+$LB results, respectively.

{\bf Fig. 6}: Plot of $\gamma(3\omega)/\gamma(0)$ (THG) as a function of 
frequency $\omega$ for helium.  The squares, open circles
and filled circles represent the LDA, the LDA$+$LB and ab-initio results
\cite{bishop1}, respectively.

{\bf Fig. 7}: Plot of $\gamma(\omega)/\gamma(0)$ (DFWM) as a function of 
frequency $\omega$ for helium.  The squares, open circles
and filled circles represent the LDA, the LDA$+$LB and ab-initio results
\cite{bishop1}, respectively.

\newpage
\begin{references}
\bibitem{gross1}E.K.U. Gross, J.F. Dobson and M. Petersilka, in {\it
Density Functional Theory}, Edited by R.F. Nalewajski, Topics in Current
Chemistry vol. 181 (Springer, Berlin, 1996).
\bibitem{casida}M. Casida, in {\it Recent Advances in Density Functional
Methods} Edited by D.P. Chong (World Scientific, Singapore, 1995).
\bibitem{zangwill}A. Zangwill and P. Soven, Phys. Rev. A {\bf 21}, 
4274 (1980).
\bibitem{stott}M.J. Stott and E. Zaremba, Phys. Rev. A {\bf 21}, 12 (1980);
Phys. Rev. A {bf 22}, E2293 (1980).
\bibitem{ghosh}S.K. Ghosh and B.M. Deb, Chem. Phys. {\bf 71}, 295 (1982);
J. Mol. Struc. (Theochem) {\bf 103}, 163 (1983).
\bibitem{barto1}L.J. Bartolotti, J. Chem. Phys. {\bf 80}, 5187 (1984);
J. Phys. Chem. {\bf 90}, 5518 (1986).
\bibitem{senatore}G. Senatore and K.R. Subbaswamy, Phys. Rev. A {\bf 35},
2440 (1987).
\bibitem{zong}H. Zong, Z. Levine and J.W. Wilkins, Phys. Rev. A {\bf 43},
4629 (1991).
\bibitem{gisberg1}S.J.A. Gisbergen, V.P. Osinga, O.V. Gritsenko, R. van 
Leeuwen, J.G. Snijders and E.J. Baerends, J. Chem. Phys. {\bf 105},
3142 (1996).
\bibitem{baner1}A. Banerjee and M.K. Harbola, Phys. Lett. A {\bf 238}, 
525 (1997).
\bibitem{gisberg2}S.J.A. Gisbergen, F. Kootstra, P.R.T. Shipper, 
O.V. Gritsenko, J.G. Snijders and E.J. Baerends, Phys. Rev. A {\bf 57},
2259 (1998).
\bibitem{gisberg3}S.J.A. Gisbergen, J.G. Snijders and E.J. Baerends, 
J. Chem. Phys. {\bf 109}, 10644 (1998).
\bibitem{baner2}A. Banerjee and M.K. Harbola, Eur. Phys. J. D {\bf 5}, 
201 (1999).
\bibitem{aiga}F. Aiga, T. Tada and R. Yoshimura, J. Chem. Phys. {\bf 111},
2878 (1999).
\bibitem{petersilka}M. Petersilka, U.J. Grossmann and E.K.U. Gross,
Phys. Rev. Lett. {\bf 76}, 1212 (1996).
\bibitem{deb1}B.M. Deb and S.K. Ghosh, J. Chem. Phys. {\bf 77}, 342 (1982).
\bibitem{barto2}L.J. Bartolotti, Phys. Rev. A {\bf 24}, 1661 (1981);
Phys. Rev. A {\bf 26}, 2243 (1982).
\bibitem{runge}E. Runge and E.K.U. Gross, Phys. Rev. Lett. {\bf 52},
997 (1984).
\bibitem{parr}R.G. Parr and W. Yang, {\it Density Functional Theory
of Atoms and Molecules} (Oxford, New-York, 1989).
\bibitem{dreizler}R.M. Dreizler and E.K.U. Gross, {\it Demsity Functional
Theory: An Approach to Many-Body Problem} (Springer Verlag, Berlin, 1990). 
\bibitem{almb} C. O. Almbladh and U. von Barth, Phys. Rev. B {\bf 31}, 3231
(1985).
\bibitem{perdew} J.P. Perdew and A. Zunger, Phys. Rev. B {\bf 23},
5048 (1981).
\bibitem{koga}T. Koga, Y. Kasai and A. J. Thakkar, Int. J. Quantum Chem.
{\bf 46}, 689 (1993).
\bibitem{van} R. van Leeuwen and E. J. Baerends, Phys. Rev. A {\bf 49}, 2421
(1994).
\bibitem{baner3} A. Banerjee and M. K. Harbola, Phys. Rev. A {\bf 60}, 3599
(1999).
\bibitem{becke}A. D. Becke, Phys. Rev. A {\bf 38}, 3098 (1988).
\bibitem{baner4}A. Banerjee, Ph. D Thesis (Devi Ahiliya Vishva Vidyalaya, 
Indore, India, 1999). 
\bibitem{gl}O. Gunnarsson and B. I. Lundquist, Phys. Rev. B {\bf 13}, 4274
(1976).
\bibitem{bishop}D. M. Bishop and B. Lam, Phys. Rev. A {\bf 37}, 464 (1988).
\bibitem{rice}J. E. Rice, J. Chem. Phys. {\bf 96}, 1992.
\bibitem{bishop1}D. M. Bishop and J. Pipin, J. Chem. Phys. {\bf 91}, 3549
(1989).
\bibitem{lehmeier}H. T. Lehemeier, W. Leupacher and A. Penzkofer, Opt.Comm.
{\bf 56}, 67 (1985).
\bibitem{new}G. H. C. New and J. F. Ward, Phys. Rev. Lett. {\bf 19}, 556 
(1967).
\bibitem{amico}I. D. Amico and G. Vignale, Phys.Rev. B {\bf 59}, 7876 (1999).
\end{references}
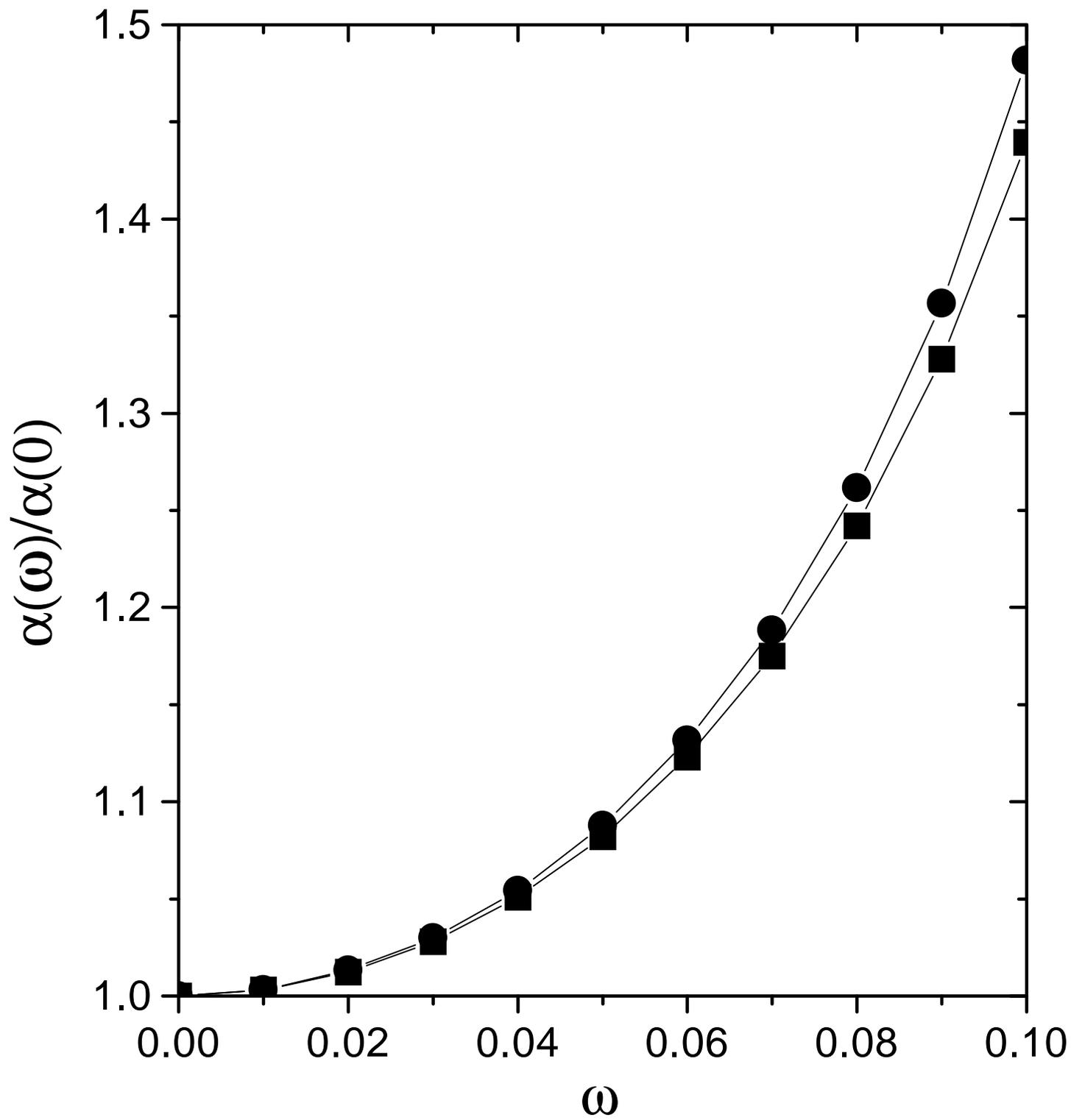
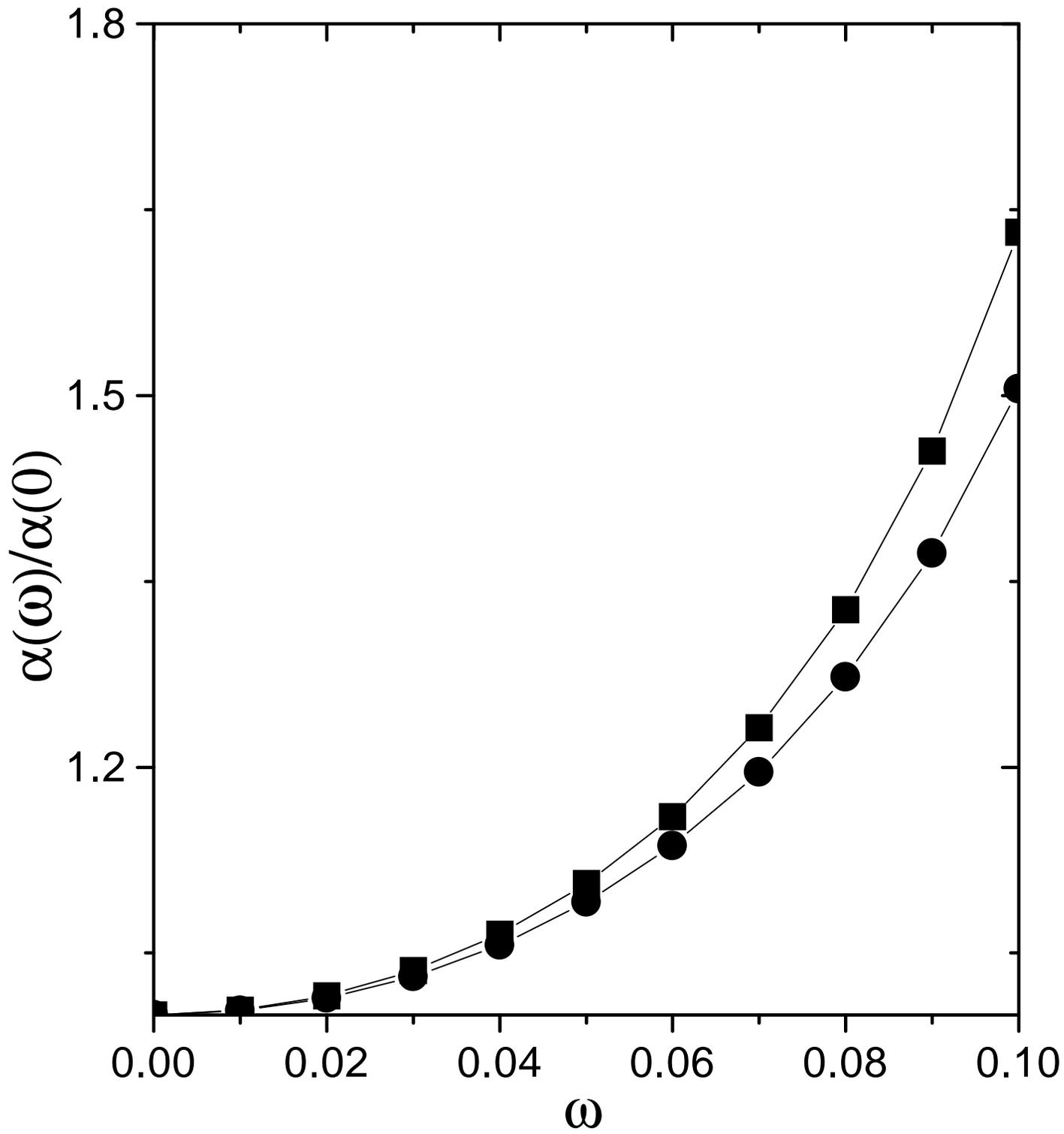
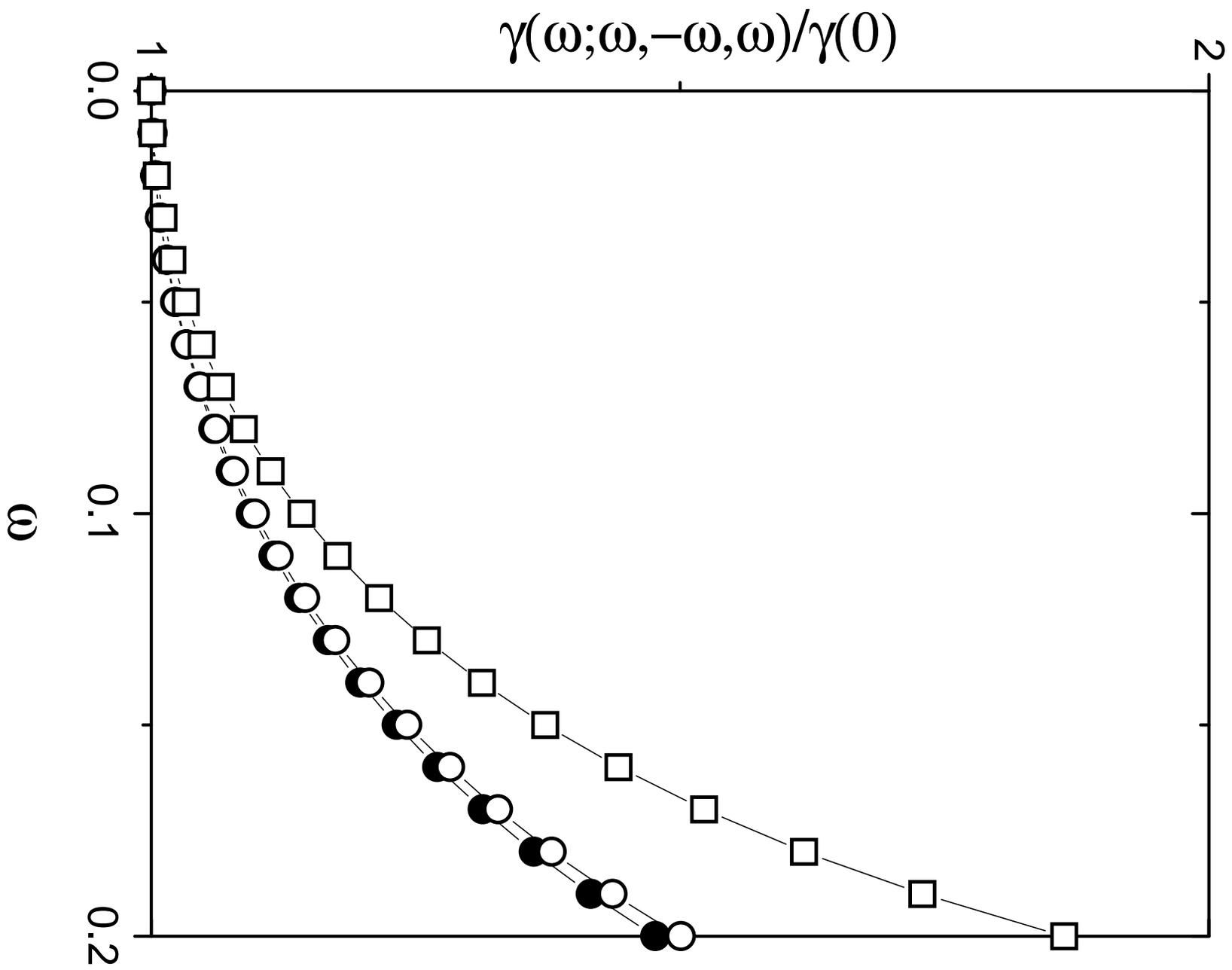
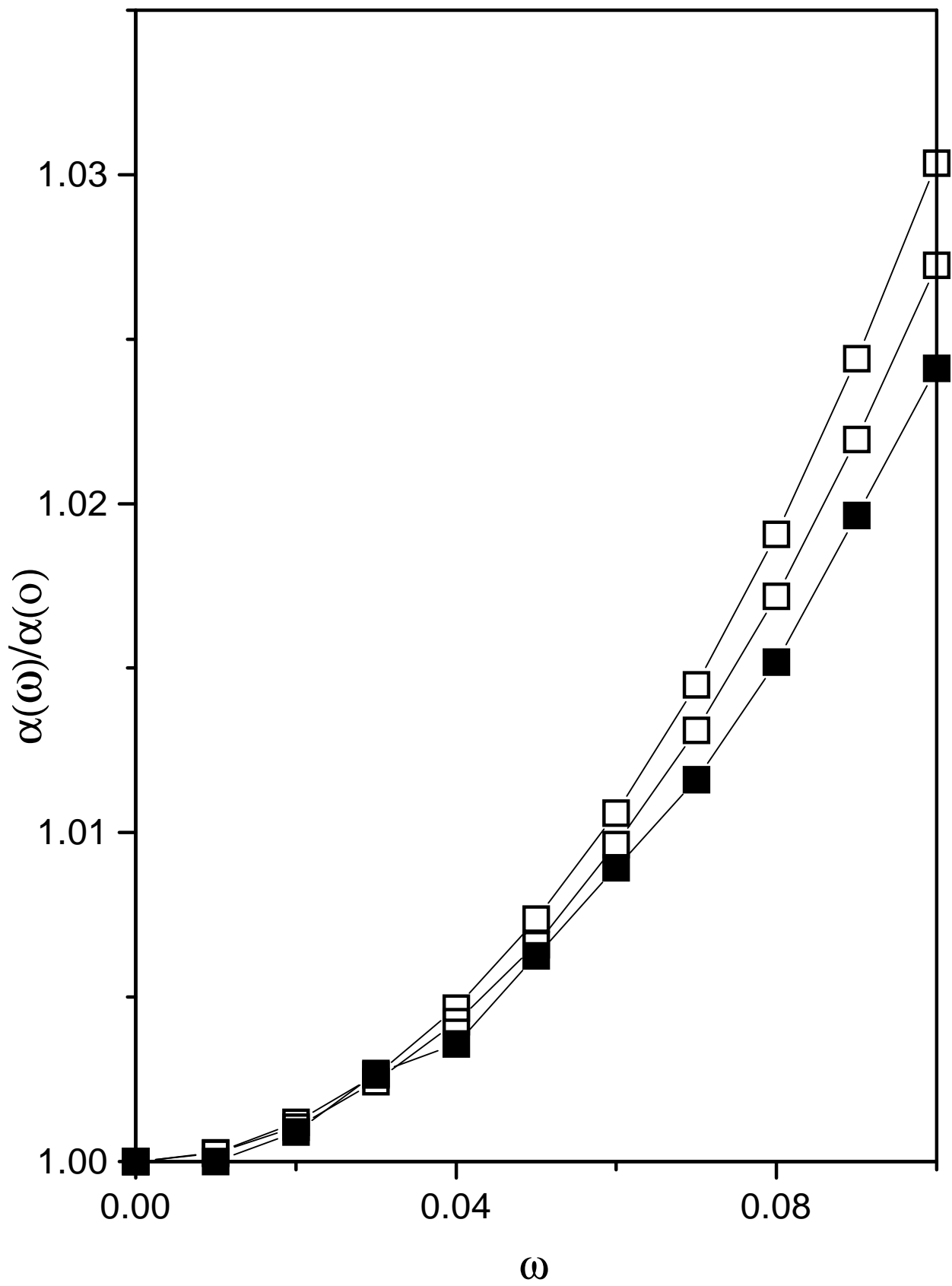
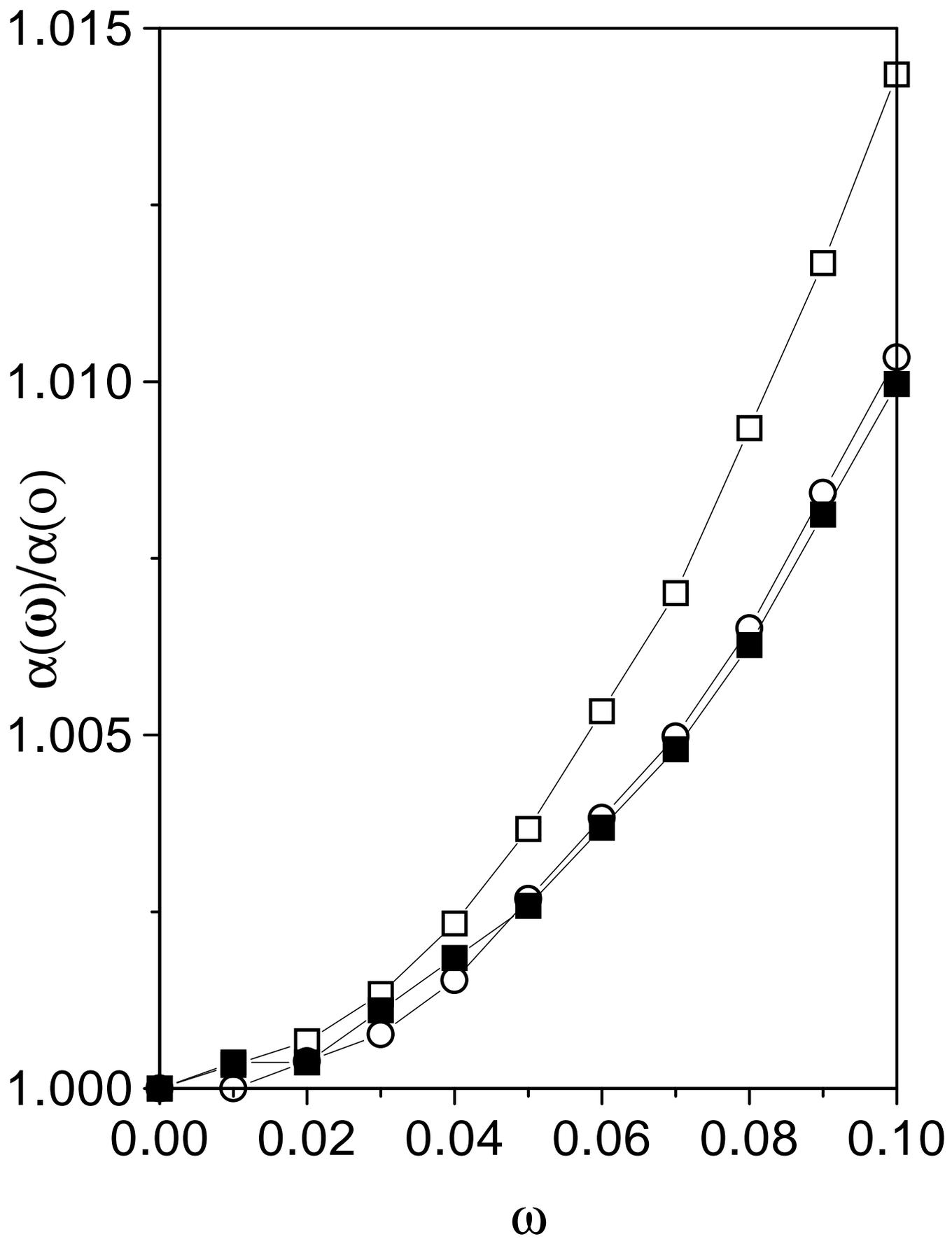
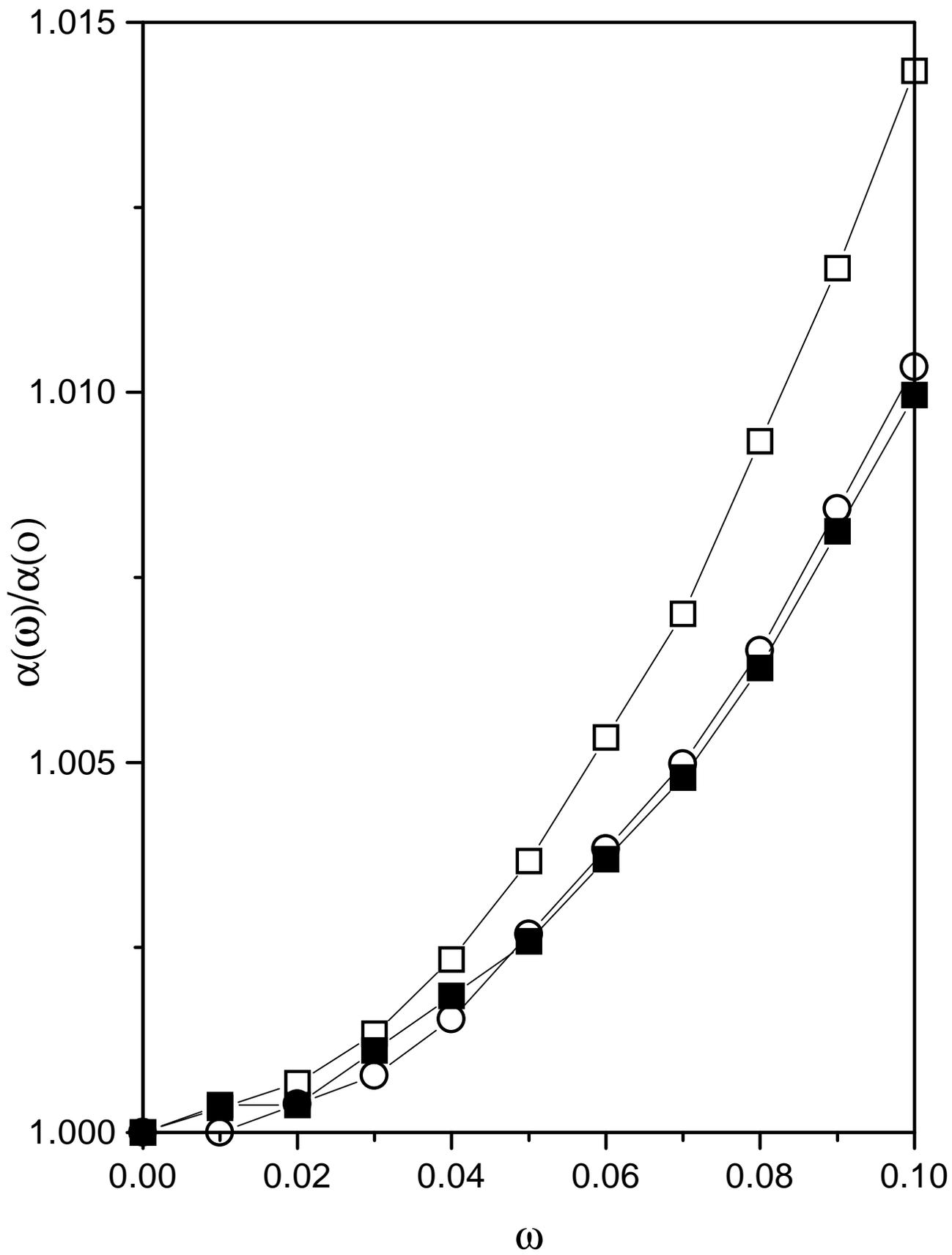
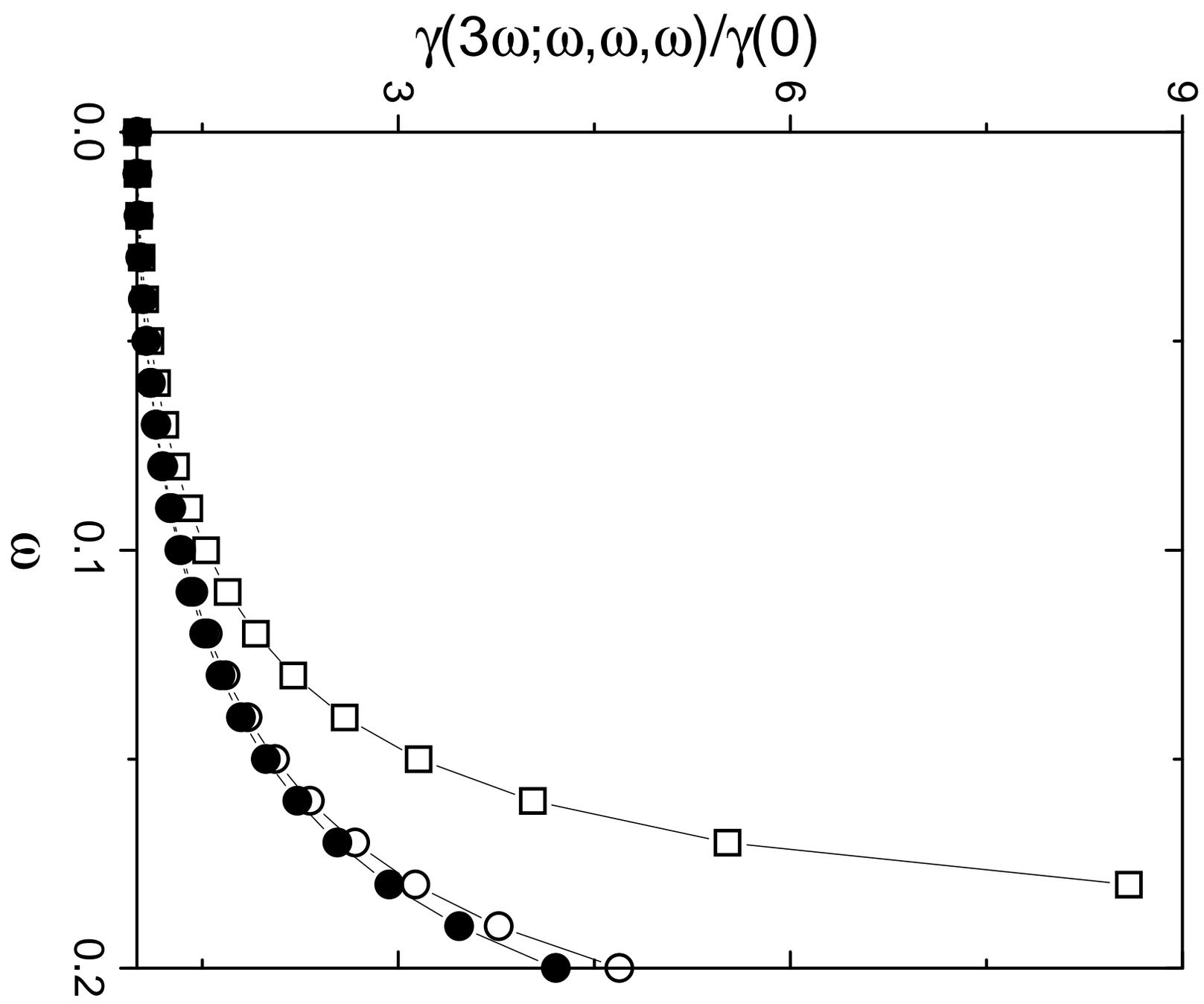
\end{document}